\documentclass[lettersize,journal]{IEEEtran}

\usepackage{amsmath,amsfonts}
\usepackage{algorithmic}
\usepackage{algorithm}
\usepackage{array}
\usepackage[caption=false,font=normalsize,labelfont=sf,textfont=sf]{subfig}
\usepackage{textcomp}
\usepackage{stfloats}
\usepackage{url}
\usepackage{verbatim}
\usepackage{graphicx}
\usepackage{cite}
\begin{document}

\title{A Bandpass Twin-T Active Filter Used in the Buchla 200 Electric Music
Box Synthesizer}

\author{Aaron D. Lanterman\thanks{School of Electrical Engineering, Georgia Institute of Technology,
777 Atlantic Drive, Atlanta, GA 30332 USA. E-mail: lanterma@ece.gatech.edu.}}

\markboth{November 2024}%
{}

\maketitle

\begin{abstract}
This paper analyzes an unusual active bandpass filter employed
in the Buchla Model 295 10 Channel Comb Filter, a synthesizer module
developed as part of the Buchla 200 Electric Music
Box by Donald Buchla. The filter consists of a peculiar rearrangement of elements
in a classic Twin-T configuration; to our knowledge, it has not been
previously addressed in the literature. As an example, we
explore its specific application in the Model 295.
\end{abstract}


\section{Introduction}

\begin{figure}[!t]
\centering
\includegraphics[width=3.5in]{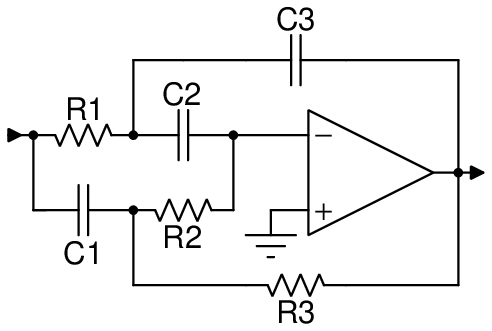}
\caption{Active bandpass filter used in the 200 Hz to 3.2 kHz
filter blocks of Figure \ref{fig:b295}.}
\label{fig:active-twint}
\end{figure}

The classic Twin-T circuit configuration is commonly used as a notch filter, in
both passive and active 
implementations 
(see \cite{wireless:parallel-t,wireless:more-parallel-t} and pp. 288-290 of
\cite{daryanani:active-network}).
Figure 6.10 on p.~192 
of \cite{huelsman:active-rc}, Figure 4(b) on 
p.~232 of \cite{wireless:parallel-t}, and Figure 1 
on p.~46 of \cite{simonton:waa-waa}
employ a Twin-T
network in the feedback loop of an inverting amplifier configuration to form a
bandpass filter. 
Figure
\ref{fig:active-twint} shows a curious rearrangement of the Twin-T
concept
that acts as a bandpass filter in the Buchla Model 295 10 
Channel Comb Filter,\footnote{Buchla used the term ``comb'' somewhat loosely,
as the bands are not equally spaced. They are instead organized along the lines
of a ``Bark'' scale \cite{zwicker:bark-scale}.} one of the modules in the Buchla 200 Electric Music Box
modular synthesizer system created by electronic music pioneer 
Donald Buchla \cite{vintage-synthesizers} in the 1970s. 

\begin{figure}[!t]
\centering
\includegraphics[width=3.2in]{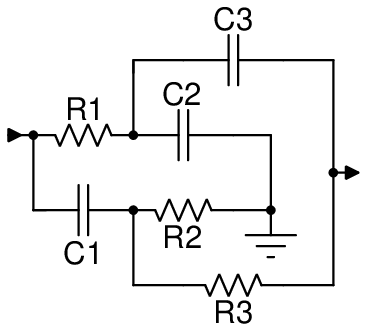}
\caption{Passive variant of the bandpass filter in 
Figure \ref{fig:passive-twint}.}
\label{fig:passive-twint}
\end{figure}

Figure \ref{fig:passive-twint} shows a passive variation, in which $C_2$ and 
$R_2$ are hooked to real ground instead of a virtual ground. If $R_1$ and $C_3$ are swapped, and
$C_1$ and $R_3$ are swapped, it then resembles the traditional passive
Twin-T notch filter; here
it is drawn slightly differently than is typical to emphasize the resemblance to
Figure
\ref{fig:active-twint}. It can also be thought of as the traditional passive
Twin-T notch filter with the ground and input connections swapped, as
illustrated in Figure 1(a) of \cite{wireless:parallel-t} (taking
$V_1=0$ and $V_2$ as the input in that figure).
A special case of this
passive filter, with $2R_1=R_2=R_3$ and $C_1/2=C_2=C_3$,
appears as Figure 8.34(a) on p.~467
of \cite{sedra-brackett:filter}, 
with pin 3 of that figure acting as the input and pin 2 acting
as the output, where it is named ``Bandpass Twin-T.'' 
Terry Watson stumbled across the same special case in his master's thesis work when
he accidentally swapped connections to a standard Twin-T network
(Figure 16 on p.~35 of \cite{watson:msthesis}, with ${\bar E}_2$ of that
figure acting as the input and ${\bar E}_1$ acting as the output). He suggests
it ``might be used as a lead-lag compensating network,'' but does not explore
it futher.

To our knowledge, this paper is the first to analyze the inverting active 
bandpass
topology of Figure
\ref{fig:active-twint}; we have been
unable to find any instances of it in the literature. A noninverting peaking
filter may be derived by swapping the input and ground 
in Figure \ref{fig:active-twint}, 
but the gain of that noninverting filter cannot drop below 
unity \cite{filter-calculator:twin-t-bandpass},
and hence it is not a strict bandpass filter.

MATLAB/Octave code
to generate the plots in this paper may be found at
\url{https://github.com/lantertronics/buchla-plots}
under the name \texttt{b295\_plots.m}.

\section{Analysis}

Straightforward nodal analysis yields the transfer function of the filter in Figure \ref{fig:active-twint}:
\begin{equation}
H(s) = -\frac
 { b_2 s^2 + b_1 s}
 {a_3 s^3 + a_2 s^2 + a_1 s + 1}, 
\label{eq:all-three-different}
\end{equation}
where
\begin{align}
 b_2 &= R_2 R_3 C_1 C_2 + R_1 R_3 C_1 (C_2 + C_3), \\
 b_1 &= (R_2+R_3) C_2 + R_3 C_1, \\
 a_3 &= R_1 R_2 R_3 C_1 C_2 C_3, \\
 a_2 &= R_1(R_2 + R_3) C_2 C_3, \\
 a_1 &= R_1 (C_2 + C_3). 
\end{align}
Although explicit expressions for the poles of 
(\ref{eq:all-three-different}) exist, they are cumbersome and provide little
insight, and an exact 
closed-form expression for the peak frequency is out of reach.
The transfer function simplifies conveniently under the special case
of $R_1 = R_3$ and $C_1 = C_3$; placing it in a canonical form with the highest
powers of the numerator and denominator having unit coefficients yields
\begin{equation}
H(s)= -\frac
 { \left[ \frac{1}{R_1 C_1} 
    + \frac{1}{R_2} \left( \frac{1}{C_1} + \frac{1}{C_2} \right) \right] 
   \left(s^2 + \frac{1}{R_1 C_1} s\right)}
 {s^3 + \frac{1}{R_1 C_1} s^2 + \frac{1}{R_1 R_2 C_1 C_2} s
      + \frac{1}{R_1^2 R_2 C_1^2 C_2}}, 
\end{equation}
which can be factored as
\begin{equation}
H(s)= -\frac
 { \left[ \frac{1}{R_1 C_1} 
    + \frac{1}{R_2} \left( \frac{1}{C_1} + \frac{1}{C_2} \right) \right] 
   \left(s + \frac{1}{R_1 C_1} s\right)} 
 {\left(s + \frac{1}{R_1 C_1}\right)
\left( s^2 + \frac{1}{R_2 C_1} s + \frac{1}{R_1 R_2 C_1 C_2} \right)}.
\label{eq:factored}
\end{equation}
The zero at $-1/(R_1 C_1)$ cancels with the corresponding pole, leaving
the classic bandpass filter form
\begin{equation}
H(s)= -\frac
 { \frac{A \omega_n}{Q} }
{ s^2 + \frac{\omega_n}{Q} s + \omega_n^2},
\label{eq:classic-bandpass}
\end{equation}
where 
\begin{align}
 A &= \frac{1}{R_2 R_1} + \left(1 + \frac{C_1}{C_2} \right), \\
 \omega_n &= \frac{1}{\sqrt{R_1 R_2 C_1 C_2}}, \label{eq:omegan} \\
 Q &= \sqrt \frac{R_2 C_1}{R_1 C_2} ,
\end{align}
with a peak at $\omega_n$, and a magnitude of $A$ at the peak.

\begin{table}[!t]
\caption{Buchla Model 295 Filterbank Part Values}
\label{tab:part-values}
\centering
\begin{tabular}{|c||c|c|c|c|}
\hline
Band & $C_1=C_3$ & $C_2$ & $R_2$ & $R'_2$\\
\hline
200 Hz & 47 nF & 10 nF & 68--168 k$\Omega$ & 89.8 k$\Omega$ \\
\hline
350 Hz & 22 nF & 4.7 nF & 91--191 k$\Omega$ & 133 k$\Omega$ \\
\hline
500 Hz & 22 nF & 2.2 nF & 91--191 k$\Omega$ & 140 k$\Omega$ \\
\hline
700 Hz & 10 nF & 1.5 nF & 150--250 k$\Omega$ & 230 k$\Omega$ \\
\hline
1000 Hz & 10 nF & 910 pF & 150--250 k$\Omega$ & 186 k$\Omega$ \\
\hline
1400 Hz & 4.7 nF & 910 pF & 150--250 k$\Omega$ & 201 k$\Omega$ \\
\hline
2000 Hz & 4.7 nF & 470 nF & 150--250 k$\Omega$ & 191 k$\Omega$ \\
\hline
3200 Hz & 2.2 nF & 470 nF & 68--168 k$\Omega$ & 133 k$\Omega$ \\
\hline
\end{tabular}
\end{table}

\begin{figure*}[!t]
\centering
\includegraphics[width=7.10in]{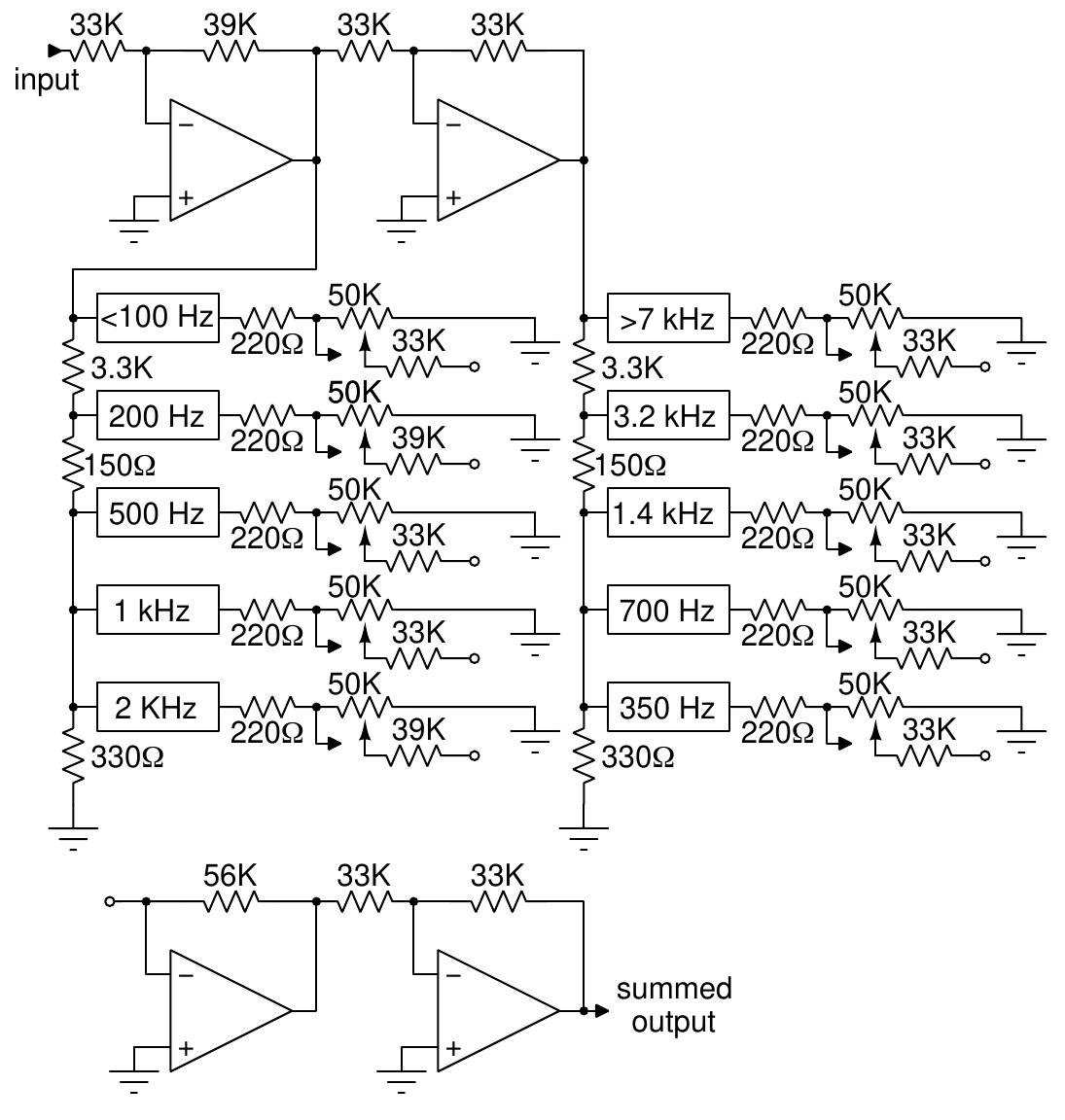}
\caption{Simplified schematic of the Buchla Model 295 10 
Channel Comb Filter. Capital K in resistance markings represents k$\Omega$
throughout the figures.  The open circles represent connections from
the 33 k$\Omega$ and 39 k$\Omega$ summing resistors to the virtual ground
of the opamp in the lower left corner. The right-facing arrows between
the 220 $\Omega$ and 50 k$\Omega$ resistances represent individual channel
outputs. The analysis in this paper assumes the opamps are ideal, so
some components meant to address nonideal opamp effects have been omitted
for clarity, namely resistors between the positive input terminals of
opamps and ground and DC blocking capacitors in series with the 33 k$\Omega$
resistors at the initial input and connecting the bottom two opamps.}
\label{fig:b295}
\end{figure*}

\section{Example}

\begin{figure}[!t]
\centering
\includegraphics[width=3.5in]{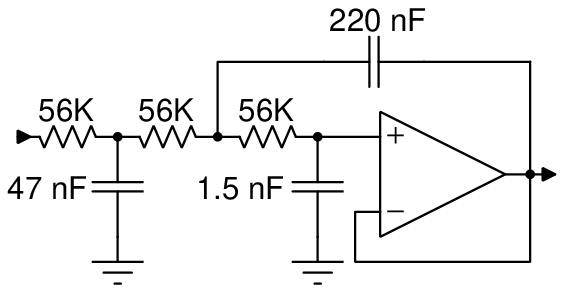}
\caption{Lowpass Sallen-Key filter for the below-100 Hz channel.}
\label{fig:lowpass_sk}
\end{figure}

\begin{figure}[!t]
\centering
\includegraphics[width=3.5in]{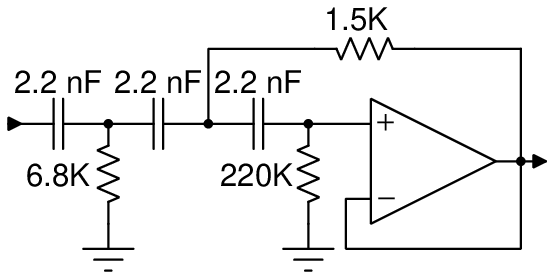}
\caption{Highpass Sallen-Key filter for the above-7 kHz channel.}
\label{fig:highpass_sk}
\end{figure}

Figure \ref{fig:b295}
presents a simplified schematic\footnote{The original schematic was found
at https://rubidium.se/\~magnus/synths/companies/buchla/Buchla\_2950\_200.jpg (last accessed December 29, 2022);
it was also found mirrored at http://fluxmonkey.com/historicBuchla/295-10chanfilt.htm.} 
of the Buchla Model 295 10 
Channel Comb Filter.\footnote{On the schematic, the designer
wrote ``trim 3.5 kc section @ 3.2 kc," so we will treat
the filter marked 3.5 kHz on the front panel and on the schematic as if it was intended to be 
3.2 kHz. One might conjecture that the designer changed their mind about
the filterbank design over the
lifetime of the product, but the front panels had already been manufactured 
to read 3.5 kHz.}
The lowest and highest channels of the filterbank consist of
three-pole Sallen-Key lowpass (Figure \ref{fig:lowpass_sk}, for frequencies less
than 100 Hz)
and highpass (Figure \ref{fig:highpass_sk}, for frequencies higher than
7 kHz) filters, 
respectively. (Since this paper assumes the opamps are ideal, elements in
the negative feedback loops of Figures \ref{fig:lowpass_sk} and
\ref{fig:highpass_sk} intended to deal with nonideal opamp effects have been
omitted for clarity.)
The middle
eight bands are implemented via Figure \ref{fig:active-twint},
with $R_1$ = 15 k$\Omega$ for 
all eight and capacitances given in Table \ref{tab:part-values}. $R_2$ for
each filter consists of a fixed resistor (68 k$\Omega$ for 200 Hz, 
91 k$\Omega$ for 350 Hz, etc.)
in series with a 100 k$\Omega$ potentiometer wired as a variable resistor, so $R_2$ in 
Table \ref{tab:part-values} is given as
a range. 
$R_3$ for each filter is a 20 k$\Omega$ potentiometer 
wired as a variable resistor.

An inverting opamp configuration at the input
multiplies the incoming signals by (39 k$\Omega$)/(33 k$\Omega)$.
Note that a second
opamp in an inverting unity gain configuration flips the signs of the input signal of 
the highpass filter and every-other bandpass filter; we will discuss the rationale for this in
Section \ref{sec:summed-output}.

The input signals for the highest and lowest channels enter their respective Sallen-Key filters directly. The input signals for the 350 Hz and 2 kHz bandpass filters are first attenuated by a voltage division factor of (330 $\Omega$)/(3.3 k$\Omega$+150 $\Omega$+330 $\Omega)$, while
the input signals for the remaining bandpass filters are first attenuated
by a voltage division factor of 
(150 $\Omega$+330 $\Omega$)/(3.3 k$\Omega$+150 $\Omega$+330 $\Omega)$. The validity of these calculations presumed that the input impedances of the various
filter stages are sufficiently high, relative to various combinations of
resistances in the attenuation ladder, that
they could be neglected.

Outputs of all ten individual filters are available to the user 
through 220 $\Omega$ short-circuit protection resistors in series with the outputs of the operational amplifiers. The individual outputs, after the 220 $\Omega$ resistors, are fed to the tops of 50 k$\Omega$ slider potentiometers, wired as volume controls with the bottoms hooked to ground. The signals at the wipers of the 
potentiometers are summed via an operational amplifier in an inverting mixer configuration, with a feedback resistance of 56 k$\Omega$. For the 200 Hz and 2 kHz channels, the resistors from the wipers to the virtual ground are
 39 k$\Omega$, whereas they are 33 k$\Omega$ for the remaining bands.
The remaining discussion will assume that the sliders are turned all the way up.
There is a slight loss in voltage
from taking the inputs to the volume controls from after
the 220 $\Omega$ instead of directly from the outputs of the filter opamps, 
but this amounts to only around 1\%, so we will neglect that effect.

\begin{figure}[thb]
\centering
\includegraphics[width=3.4in]{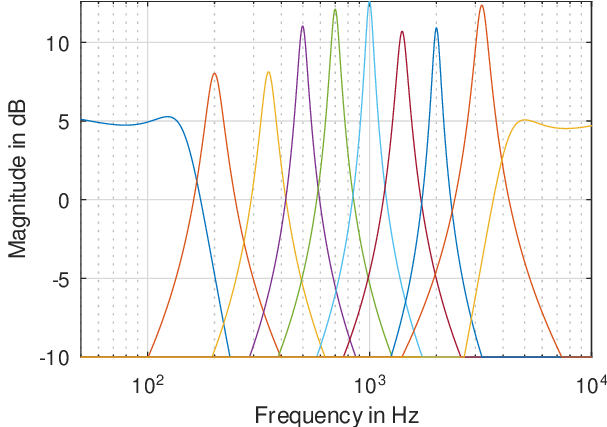}
\caption{Frequency responses of the 10 channels assuming $R_2=R'_2$ and
$R_3=R_1$.}
\label{fig:initial-responses}
\end{figure}

\begin{figure}[thb]
\centering
\includegraphics[width=3.4in]{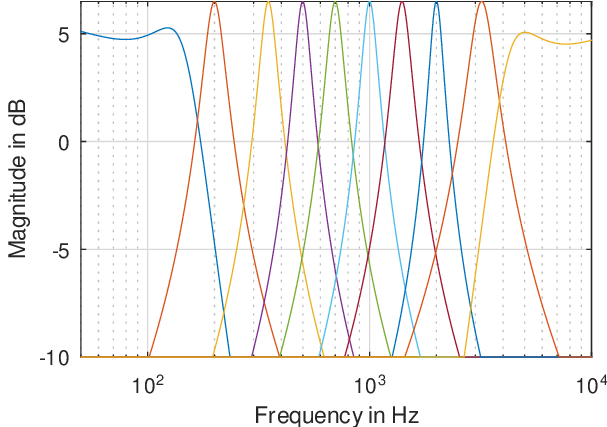}
\caption{Frequency responses of the 10 channels assuming the 
$R_2$ and $R_3$ values specified in Table \ref{tab:calibration-results}.}
\label{fig:calibrated-responses}
\end{figure}

\subsection{Calibration}

As an initial estimate for choosing $R_2$,
if we assume $R_3=R_1=15$ k$\Omega$, we could use (\ref{eq:omegan}) to 
compute a candidate
value
according to the target peak frequency $f_c$ given in the left column of
Table \ref{tab:part-values}:
$R'_2 = 1/ (2 \pi f_c R_1 C_1 C_2)$.
Comparing the right two columns of Table \ref{tab:part-values}, we see that
the computed $R'_2$ values are within the range allowed by the potentiometer.
Letting $R_2=R'_2$ yields the magnitude frequency responses shown in
Figure \ref{fig:initial-responses}, using equations from p.~30 of
\cite{leach:filterpot} for the Sallen-Key filters. 

On the Buchla Model 295 schematic, the designer 
wrote instructions to trim each channel to achieve 6.5
dB of gain at the target frequency with the sliders set to maximum volume. 
They
also wrote that the ``100K trimmers tune frequencies,'' and the 
``20K trimmers tune
Qs (and gains).'' Unfortunately, it is not so simple, as all of those quantities
are effected by both trimmers. We iteratively adjusted the potentiometers
by alternately changing the 20 k$\Omega$
 trimmer to achieve the desired gain and 
the 100 k$\Omega$
 trimmer to achieve the desired peak frequency, until the peaks were
within 1 Hz and 0.01 dB of the desired targets. The resulting $R_2$ and $R_3$
values
are given in Table \ref{tab:calibration-results}, with the corresponding 
frequency responses illustrated in Figure \ref{fig:calibrated-responses}.

\begin{table}[!t]
\caption{Buchla Model 295 Filterbank Calibration Results}
\label{tab:calibration-results}
\centering
\begin{tabular}{|c||c|c|c|c|}
\hline
Band & $R_2$ & $R_3$ & Q\\
\hline
200 Hz & 93 K$\Omega$ & 13.98 k$\Omega$ & 4.522 \\
\hline
350 Hz & 136.6 k$\Omega$ & 14.04 k$\Omega$ & 5.395  \\
\hline
500 Hz & 149.5 k$\Omega$ & 13.15 k$\Omega$ & 5.941 \\
\hline
700 Hz & 242 k$\Omega$ & 12.41 k$\Omega$ & 5.422  \\
\hline
1000 Hz & 199.5 k$\Omega$ & 12.75 k$\Omega$ & 5.989 \\
\hline
1400 Hz & 209.8 k$\Omega$ & 12.74 k$\Omega$ & 5.222  \\
\hline
2000 Hz & 200.5 k$\Omega$ & 13.4 k$\Omega$ & 6.946  \\
\hline
3200 Hz & 171.2 k$\Omega$ & 11.3 k$\Omega$ & 3.721  \\
\hline
\end{tabular}
\end{table}

If the denominators
of the resulting transfer functions are factored as $(s+a)[s+(\omega_n/Q)+\omega_n^2]$,
where $a$ is real, we find $\omega_n/(2 \pi)$ values within 1 Hz of the
target frequencies. This suggests that the complex poles and the zero at the origin dominate the response, even though the real pole does not exactly cancel with the negative real zero as in (\ref{eq:factored}). Table \ref{tab:calibration-results} lists the corresponding Q values.

\subsection{Summed Output}
\label{sec:summed-output}

\begin{figure}[!t]
\centering
\includegraphics[width=3.4in]{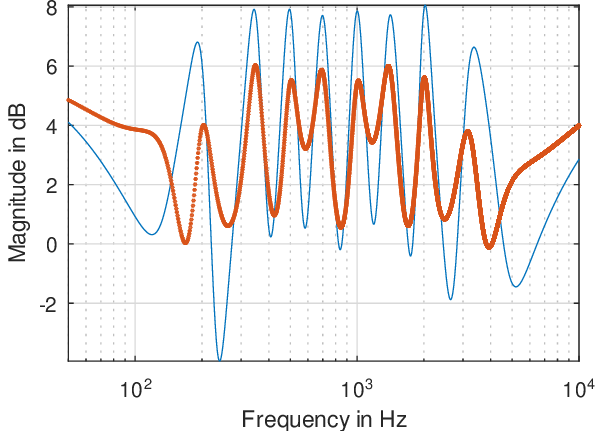}
\caption{Frequency response of the summed output of the Buchla Model 295. The thin
solid line corresponds to omitting the inversion of certain channels. 
The thick dotted line includes the inversions (although the dots are spaced
close enough that they appear solid throughout most of the graph). }
\label{fig:summed-output}
\end{figure}

Figure \ref{fig:summed-output} shows the summed output of all 10 channels,
with all sliders at full volume. 
The thin solid line illustrates what happens if we ignore the opamp in the
upper right corner of Figure \ref{fig:b295}, which inverts the highpass channel and
every-other bandpass
channel. The thick dotted line shows the result when we include those
counterintuitive inversions,
and illustrates why they are there. The phase responses of the neighboring
filters naturally conspire to produce some wide amplitude variations, which
are somewhat smoothed by inverting some of the channels. The resulting frequency
response is still not particularly flat; the Buchla Model 295 is a musical effect,
not a precision equalizer. 

\section{Conclusion}
We do not know, at present, where else the topology of 
Figure \ref{fig:active-twint} may have been used, or why the designer of
the Model 295 chose it over the cascaded three-pole Sallen-Key filters used
in the Mosdel 294 4 Channel Comb Filter or the cascaded multiple feedback bandpass filters employed in the 16-band Model 296 Programmable Spectral Processor.
For that matter, it is unclear whether the filter truly originated with Donald Buchla (assuming he was the designer, which is likely),
or may yet be found in a ``quaint and curious volume of forgotten lore,''
as penned by Edgar Allen Poe,
despite
our thorough literature search.

Future work could explore what advantages and disadvantages Figure \ref{fig:active-twint} has over other bandpass filter topologies, and formulate
design procedures for the filter. We were unable to reverse engineer 
the designer's
thought process in choosing component values; issues like the choice of 
39 k$\Omega$ instead of 33 k$\Omega$ for two of the filters are suggestive of
some trial-and-error salting-to-taste on the designer's part.

\section*{Acknowledgments}
We thank Neil Johnson for finding passive version of the filter in 
\cite{sedra-brackett:filter,watson:msthesis}, Peter Singfield
for finding the articles in 
Wireless World \cite{wireless:parallel-t,wireless:more-parallel-t} and his
insights about the noninverting variation of the filter,
and Don Tillman for finding the article in 
Popular Electronics \cite{simonton:waa-waa} and his help in
developing intuition about the Buchla Model 295,
particularly the role of the inversion
of some of the channels. We particularly thank Magnus Danielson for his detailed
reading of a draft of this paper and his comments.

\newpage

\bibliographystyle{IEEEtran}
\bibliography{twint_arxiv}

\end{document}